\documentclass[amsmath,amssymb,aps,twocolumn]{revtex4-2}

\usepackage{graphicx}
\usepackage{dcolumn}
\usepackage{bm}

\usepackage{braket}
\usepackage{xcolor}
\usepackage{mathtools}

\usepackage[colorlinks=true,citecolor=blue,linkcolor=blue,urlcolor=blue]{hyperref}

\newcommand{\abs}[1]{\left\vert#1\right\vert}
\newcommand{\defeq}{\vcentcolon=}

\begin{document}


\title{Refined Tsirelson Bounds on Multipartite Bell Inequalities}

\author{Rain Lenny}
\author{Dana Ben Porath}
\author{Eliahu Cohen}%
 \email{eliahu.cohen@biu.ac.il}
\affiliation{Faculty of Engineering and the Institute of Nanotechnology and Advanced Materials, Bar Ilan University, 5290002 Ramat Gan, Israel
}%

\date{\today}

\begin{abstract}
Despite their importance, there is an on-going challenge characterizing multipartite quantum correlations. The Svetlichny and Mermin-Klyshko (MK) inequalities present constraints on correlations in multipartite systems, a violation of which allows to classify the correlations by using the non-separability property. In this work we present refined Tsirelson (quantum) bounds on these inequalities, derived from inequalities stemming from a fundamental constraint, tightly akin to quantum uncertainty. Unlike the original, known inequalities, our bounds do not consist of a single constant point but rather depend on correlations in specific subsystems (being local correlations for our bounds on the Svetlichny operators and bipartite correlations for our bounds on the MK operators). 
We analyze concrete examples in which our bounds are strictly tighter than the known bounds.

\end{abstract}

\maketitle

\section{Introduction}
In their seminal work \cite{einstein1935can}, Einstein, Podolsky and Rosen (EPR) questioned the completeness of quantum mechanics, highlighting the fundamental implications of entanglement. They suggested that supplementing quantum mechanics with, what would later be called, local hidden variables (LHV) could mend nonlocality. Bell demonstrated that this could not be. He proposed an inequality that can be experimentally tested, a violation of which proves that no LHV model can reproduce measurement outcomes prescribed by quantum theory \cite{bell1964einstein}. Building upon Bell's work, Clauser, Horne, Shimony and Holt (CHSH) proposed another inequality to test LHV theories \cite{clauser1969proposed}, which states that an LHV model describing a bipartite system of parties $A^{(1)}$ and $A^{(2)}$, each with two observables, must satisfy: $\abs{\braket{A^{(1)}_0A^{(2)}_0} + \braket{A^{(1)}_0A^{(2)}_1} + \braket{A^{(1)}_1A^{(2)}_0} - \braket{A^{(1)}_1A^{(2)}_1}} \leq 2$. A violation of this inequality implies that the system cannot be described by an LHV model \cite{brunner2014bell,lambare2022meaning}. Bell inequalities, such as the CHSH, have become a central focus of research in quantum information science, loopholes in their realization were closed and violations were recorded in numerous experiments \cite{tittel1998violation,rowe2001experimental,hasegawa2003violation,lima2010experimental,dressel2014avoiding}.

Along with the major advances in quantum information science, new inequalities for correlations in multipartite ($N \geq 3$) systems were derived, most notably by Svetlichny \cite{svetlichny1987distinguishing}, Mermin \cite{mermin1990extreme}, Belinskiĭ and Klyshko \cite{belinskiĭ1993interference}. The ``classical'' bound on the Svetlichny operators is given under an LHV model exhibiting an arbitrary partial separability (Eq. (1) in \cite{seevinck2002bell}). Thus, a violation of this bound implies that the system does not exhibit any partial separability, i.e., it is genuinely non-separable. Similarly, the classical bound on the Mermin-Klyshko (MK) operators is given under an LHV model exhibiting full separability of all the parties (Eq. (1) in \cite{collins2002bell}). Thus, a violation of this bound indicates that the system does not exhibit full separability, i.e., it is non-fully-separable. Violations of these multipartite Bell inequalities were also recorded in experiments \cite{pan2000experimental,zhao2003experimental,lavoie2009experimental,alsina2016experimental}.

Several recent works \cite{carmi2018significance,carmi2019bounds,carmi2019relativistic,hofmann2019local} have emphasized the significance of local uncertainty relations and the interplay between local and nonlocal correlations in determining the extent of quantum nonlocality. This understanding has also given rise to various bounds on quantum correlations \cite{carmi2019bounds,ben2023leggett}, as well as entanglement detection criteria \cite{peled2021correlation,lenny2023multipartite}. In this work we generalize and apply this notion to arbitrary $N$-party systems in order to derive refined Tsirelson (quantum) bounds on the $N$-party Svetlichny \cite{seevinck2002bell} and MK inequalities \cite{collins2002bell}. Mathematically, our inequalities stem from the  semidefinite positiveness of the covariance matrix, which physically corresponds to the ``relativistic independence'' principle \cite{carmi2019relativistic}. Both the Svetlichny and MK operators are comprised of nonlocal correlations between \emph{all} the parties. In our inequalities, the bounds depend on specific local correlations evaluated in each party, i.e., $\braket{A^{(n)}_0A^{(n)}_1}$, for
our bounds on the Svetlichny operators; and bipartite correlations evaluated between each two parties, i.e., $\braket{A^{(n)}_0A^{(m)}_1}$, for our bounds on the MK operators (for an odd number of parties). We provide concrete examples in which our bounds are not only strictly tighter than the known bounds but also demonstrate cases where our quantum bounds coincide with the classical bounds. In addition to the clear implication of tighter quantum bounds, which make it easier to discern between quantum correlations and those originating from post-quantum models, such as Popescu-Rohrlich boxes \cite{popescu1994quantum}, we interpret our results as complementarity relations setting the interplay between multipartite nonlocal correlations and lower-order correlations (local or bipartite correlations depending on the specific inequality).

For the upcoming analysis, we consider an $N$-party system  $\{A^{(n)}\}_{n=1}^{N}$, wherein each party may measure two operators $A_0^{(n)}$ and $A_1^{(n)}$, with measurement outcomes being $\pm 1$. Furthermore, for measurements on two or more systems, such as $A_0\otimes A^{(2)}_0$, we now dispose of the tensor product sign, writing $A_0 A^{(2)}_0$ in accordance with \cite{collins2002bell,seevinck2002bell}. 

The rest of the paper is organized as follows. In Sec. \ref{ineq_from_covariance} we introduce the inequalities used to derive our results. We derive our refined Tsirelson bounds on the Svetlichny operators in Sec. \ref{tighter_bounds_svetlichny} and on the MK operators in Sec. \ref{tighter_bounds_mk}, examples of which are given in Sec. \ref{SvetExample} and \ref{MermExample} respectively. We discuss a relation between our bounds and the corresponding algebraic bounds in Sec. \ref{algebraic_bounds} and conclude in Sec. \ref{Conclusions}.

\section{Inequalities from the covariance matrix}\label{ineq_from_covariance}

We may reduce a multipartite system $\{A^{(n)}\}_{n=1}^{N}$ into a bipartite one by joining different parties, referring to one part as a single party $X$, and the other part by $Y$ as follows,
\begin{equation}
\underbrace{A^{(1)}A^{(2)} \ldots A^{(P)}}_{X} \underbrace{A^{(P+1)} \ldots A^{(N)}}_{Y}.
\end{equation}
Using the semidefinite positiveness of the covariance matrix when it is associated with operators, such as in quantum mechanics, the following inequalities can be derived: 
\begin{equation}\label{MainIneqXY}
\abs{\braket{X_iY_k} \pm \braket{X_jY_k}} \leq \sqrt{ 2 \pm  \braket{\{X_i,X_j\}}},
\end{equation}
\begin{equation}\label{MainIneqYX}
\abs{\braket{X_kY_i} \pm \braket{X_kY_j}} \leq \sqrt{ 2 \pm  \braket{\{Y_i,Y_j\}}},
\end{equation}
wherein $X_iY_k$ is an operator acting on the entire system (see Theorem 1 in \cite{carmi2018significance}, as well as Appendix \ref{CovarianceProofs} here). 

\section{Refined Tsirelson bounds on Svetlichny operators}\label{tighter_bounds_svetlichny}
\subsection{Svetlichny operators}
The Svetlichny operators can be recursively defined as (Eq. (10) in \cite{seevinck2002bell}):
\begin{equation}\label{SvetRecur}
\mathcal{S}_N^{\pm} = \mathcal{S}_{N-1}^{\pm}A_0^{(N)} \mp \mathcal{S}_{N-1}^{\mp}A_1^{(N)}.
\end{equation}
The initial case is the CHSH operator: $\mathcal{S}_2^- = \left(\mathcal{S}_2^{+}\right)^{'} = A^{(1)}_0A^{(2)}_0 + A^{(1)}_0A^{(2)}_1 + A^{(1)}_1A^{(2)}_0 - A^{(1)}_1A^{(2)}_1$, wherein $'$ denotes the act of interchanging the labels of the operators, e.g., $(A_0B_1C_0)^{'} = A_1B_0C_1$.

\subsection{Refined Tsirelson bounds}
Consider a bipartite system of parties $A^{(1)}$ and $A^{(2)}$. From Eq. \eqref{MainIneqXY}, combined with the triangle inequality, we obtain:
\begin{equation}\label{S2MinusBound1}
\begin{split}
\abs{\braket{\mathcal{S}_2^-}} \leq \sqrt{ 2 + \braket{\{A^{(1)}_0,A^{(1)}_1\}}} \\ 
+ \sqrt{ 2 - \braket{\{A^{(1)}_0,A^{(1)}_1\}}},
\end{split}
\end{equation}
in which the bound can be simplified to be:
\begin{equation}\label{TransitionEq}
   2\sqrt{1+\sqrt{1-\left( \frac{1}{2} \braket{\{A^{(1)}_0,A^{(1)}_1\}} \right)^2}}.
\end{equation}
In a similar manner, but using Eq. \eqref{MainIneqYX}, we obtain:
\begin{equation}\label{S2MinusBound2}
\abs{\braket{\mathcal{S}_2^-}} \leq 2\sqrt{1+\sqrt{1-\left( \frac{1}{2} \braket{\{A^{(2)}_0,A^{(2)}_1\}} \right)^2}}.
\end{equation}
These are known inequalities, which are presented in Refs. \cite{carmi2018significance,carmi2019bounds}. It is easy to see that similar bounds can be derived for $\mathcal{S}_2^+$.

By the recursive definition of the Svetlichny operators (Eq. \eqref{SvetRecur}), our bounds on $\abs{\braket{\mathcal{S}_{N}^{\pm}}}$ are twice the bounds on $\abs{\braket{\mathcal{S}_{N-1}^{\pm}}}$ and we have the following result for every $N$-party Svetlichny operator in the quantum regime:
\begin{equation}\label{BoundsSvetGen}
\abs{\braket{\mathcal{S}_N^{\pm}}} \leq 2^{N-1} \sqrt{1+\sqrt{1-\eta^{(n)}}},
\end{equation}
wherein $ \eta^{(n)} \defeq \left(\frac{1}{2} \braket{\{A_0^{(n)},A_1^{(n)}\}}\right)^2$, for $n\in \{1,2,\ldots,N\}$. Note that $0 \leq \eta^{(n)} \leq 1$.

It is easy to see that our bounds are local, in the sense that $\eta^{(n)}$ represents correlations between two $A^{(n)}$ observables. In contrast, the Svetlichny operator is comprised of multipartite correlations between all the parties. The fundamental consequence of this observation is that local correlations limit our ability to detect nonlocal correlations (in this case genuine non-separability) and that there is a complementary relation between them.

The maximal value of our bound is $2^{N-1}\sqrt{2}$, which is the known Tsirelson bound \cite{cirel1980quantum,seevinck2002bell} (the quantum bound), obtained for the minimal value of local correlations, i.e., when $\eta^{(n)} = 0$. The minimal value of our bound is $2^{N-1}$, the known classical bound for an LHV model factorizable for an arbitrary bipartition (Eq. (1) in \cite{seevinck2002bell}), and is obtained for the maximal value of local correlations, i.e., when  $\eta^{(n)} = 1$. Thus, our bounds are tighter Tsirelson bounds on the Svetlichny operator. We emphasize that if a single party has maximal local correlations, $\eta^{(n)} = 1$, the Svetlichny operator cannot cross the classical bound, and genuine non-separability cannot be detected. This underlies our complementarity relation, demonstrated in the next section.

\section{Examples - The Svetlichny operator}\label{SvetExample}
In the following examples, we test our refined bounds on the Svetlichny operator $\mathcal{S}_3^-$, demonstrating the aforementioned complementarity relation. We do so by considering the tripartite Greenberger, Horne and Zeilinger (GHZ) state \cite{greenberger1989going,mermin1990quantum}:  
\begin{equation}
    \ket{\text{GHZ}} = \frac{\ket{000}+\ket{111}}{\sqrt{2}}.
\end{equation}
The GHZ state is genuinely entangled and saturates the known Tsirelson bound on the Svetlichny operators \cite{seevinck2002bell}. We will use it, with a meticulous choice of the parties' measurement operators, to demonstrate our results. Using states which are not as entangled as the GHZ state might not give such drastic differences between our bounds and the known bounds, but the interpretation given by our bounds, as will be further explored via the example, nevertheless holds.

We take the parties' measurement operators to be \cite{seevinck2002bell}:
\begin{equation}\label{ExampleOps}
    A^{(n)}_i = \cos\left(\theta_i^{(n)}\right)\sigma_X + \sin\left(\theta_i^{(n)}\right)\sigma_Y,
\end{equation}
where $i \in \{0,1\}$. These operators return measurement outcomes of $\pm 1$, corresponding to their two eigenvalues.

\subsection{Saturating the known Tsirelson bound}

Taking the phases of the operators to be
$(\theta_0^{(1)},\theta_0^{(2)},\theta_0^{(3)}) = (-\pi/4,0,0)$ and $(\theta_1^{(1)},\theta_1^{(2)},\theta_1^{(3)}) = (\pi/4,\pi/2,\pi/2)$, yields saturation of the known Tsirelson bound:
\begin{equation}
   \abs{\braket{\mathcal{S}_3^{-}}}=  4 \sqrt{1+\sqrt{1-\eta^{(n)}}} = 4\sqrt{2},
\end{equation}
for $n \in \{1,2,3\}$. As stated earlier, in this case \emph{all} the local correlations are zero, $\eta^{(n)} = 0$, allowing the Svetlichny operator to reach its maximal value.

\begin{figure}[ht]
\includegraphics[scale=0.53]{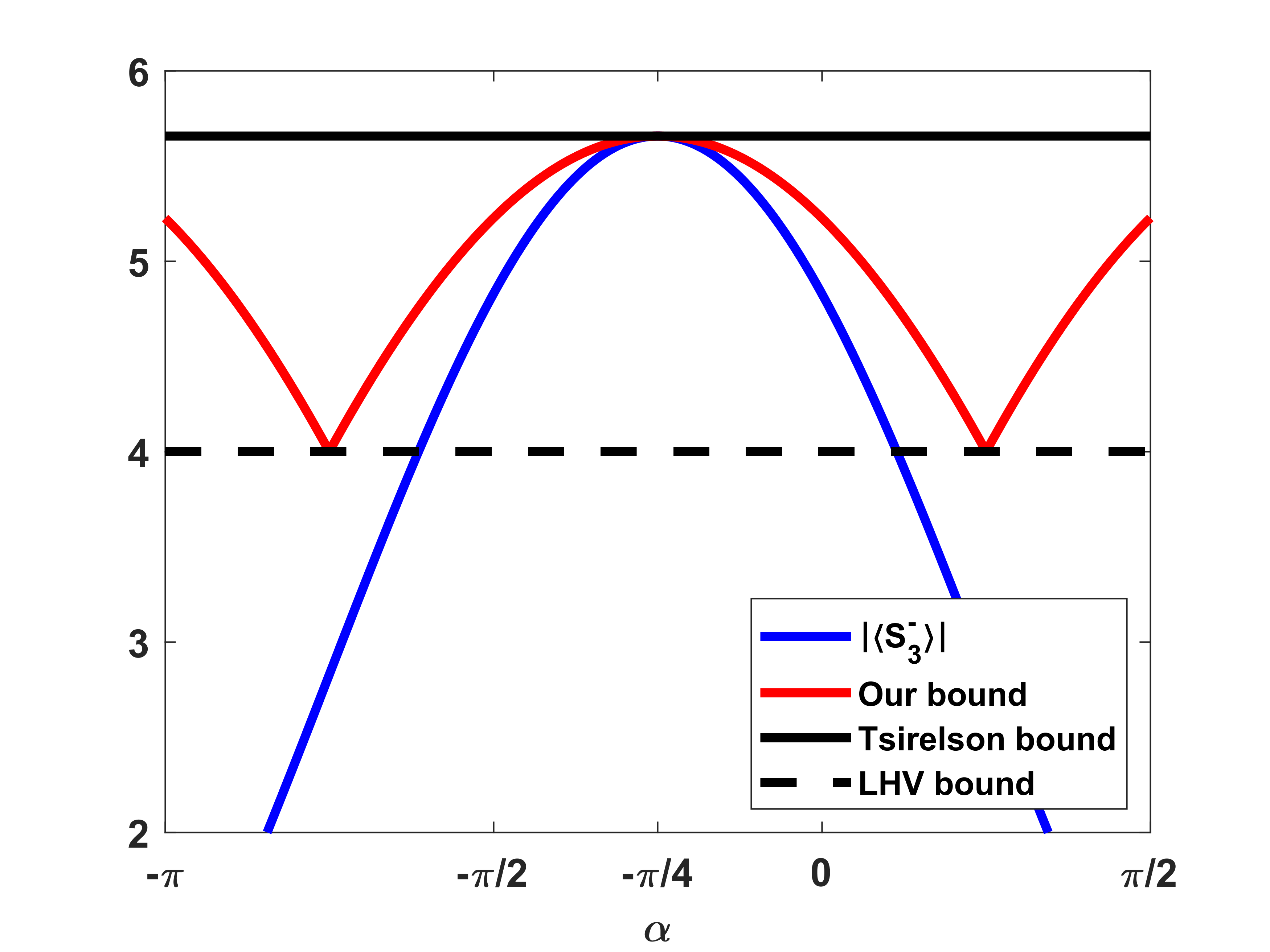}
\caption{\textbf{Demonstrating the bounds from Eq. \eqref{BoundsSvetGen}.} Plotted are the Svetlichny operator $\mathcal{S}_3^-$ and our bound, calculated for a tripartite GHZ state and the operators of Eq. \eqref{ExampleOps} with $(\theta_0^{(1)},\theta_0^{(2)},\theta_0^{(3)}) = (\alpha,0,0)$ and $(\theta_1^{(1)},\theta_1^{(2)},\theta_1^{(3)}) = (\pi/4,\pi/2,\pi/2)$. It can be seen that for the chosen operators, our new bound constitutes a stricter quantum constriction on the Svetlichny operator.}
\centering
\label{GHZstrictTight}
\end{figure}

Furthermore, changing the phase of $A_0^{(1)}$ using the parameter $\alpha$, such that $(\theta_0^{(1)},\theta_0^{(2)},\theta_0^{(3)}) = (\alpha,0,0)$ and $(\theta_1^{(1)},\theta_1^{(2)},\theta_1^{(3)}) = (\pi/4,\pi/2,\pi/2)$, proves to be an important example of a case in which our bounds are strictly tighter than the known Tsirelson bound. Our results are plotted in Fig. \ref{GHZstrictTight}, in which our bound, plotted in the figure in red, is given by $4 \sqrt{1+\sqrt{1-\eta^{(1)}}}$ (Eq. \eqref{BoundsSvetGen}). For $\alpha =-\pi/4  $ the previously mentioned saturation of the known Tsirelson bound can be observed.

\subsection{Demonstrating the complementarity between local and nonlocal correlations}

Here, we place the following phases, using the parameter $\alpha$, in the operators from Eq. \eqref{ExampleOps}:
$(\theta_0^{(1)},\theta_0^{(2)},\theta_0^{(3)}) = (0,0,0),$ $(\theta_1^{(1)},\theta_1^{(2)},\theta_1^{(3)}) = (\alpha,\alpha,\alpha)$.

\begin{figure}[t]
\includegraphics[scale=0.53]{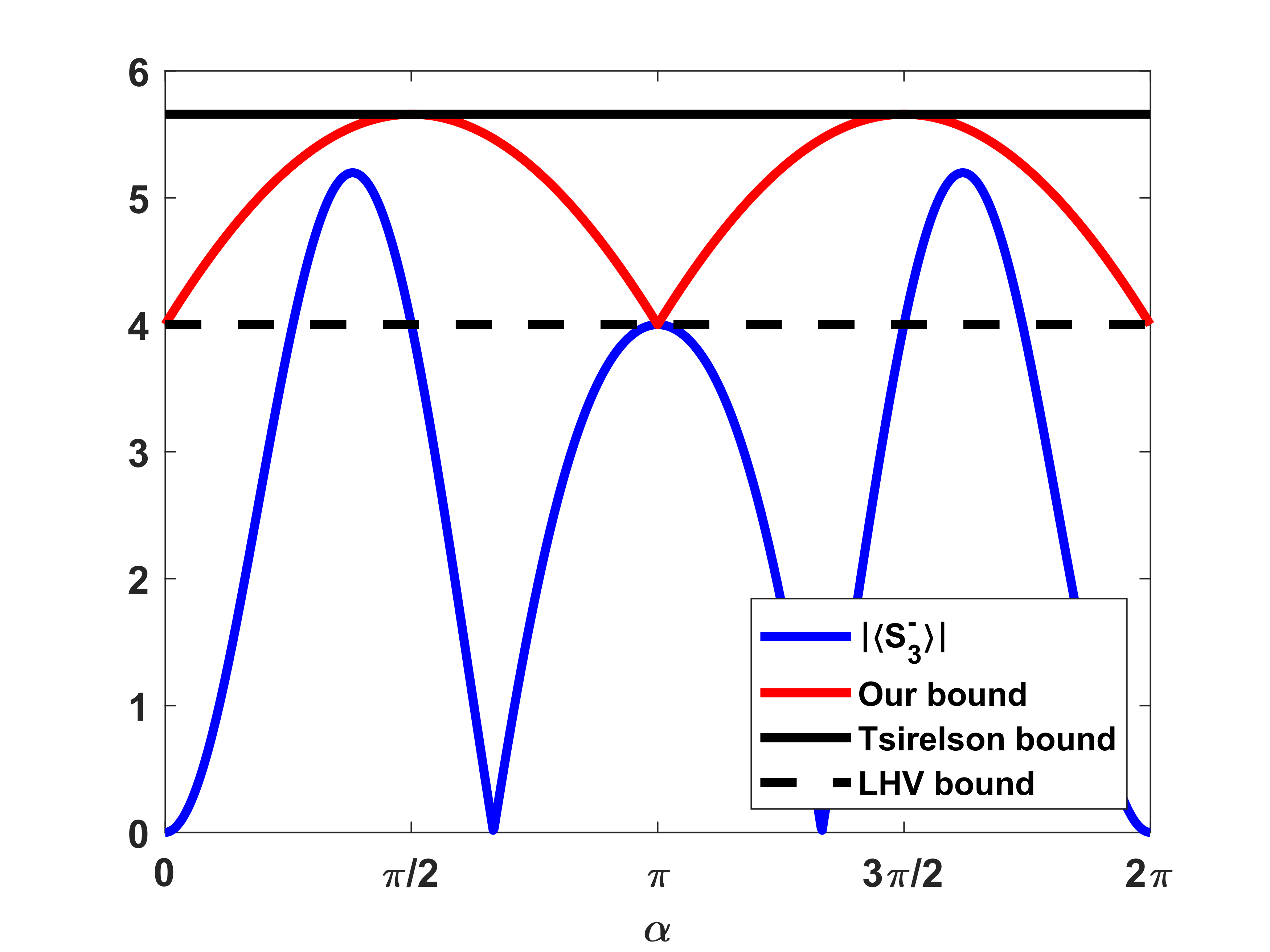}
\caption{\textbf{Demonstrating the complementarity of local and nonlocal correlations.} Plotted are the Svetlichny operator $\mathcal{S}_3^-$ and our bounds, calculated for a tripartite GHZ state. The known Tsirelson bound is $4\sqrt{2}$ in the tripartite case. The complementarity between local and nonlocal correlation is demonstrated, and in particular, for $\alpha = \pi$ our bound is equal to $4$ and thus genuine non-separability cannot be detected.}
\centering
\label{PolSimGHZ}
\end{figure}

Presented in Fig. \ref{PolSimGHZ} are the Svetlichny operator and our bound (the three bounds given by Eq. \eqref{BoundsSvetGen} are identical in this case). 
It can be seen that for $\alpha = \pi$ we have a saturation of the classical bound:
\begin{equation}
\abs{\braket{\mathcal{S}_3^-}} = 4 \sqrt{1+\sqrt{1-\eta^{(n)}}} = 4,
\end{equation}
for $n \in \{1,2,3\}$. For this case of $\alpha = \pi$, the operators are $A_0^{(n)} = -A_1^{(n)} = \sigma_X$ (for all $n$), yielding a maximal value of the local correlations $\eta^{(n)} = 1$. Since the local correlations are maximal, our bound is equal to its minimal value, which is the classical bound, prohibiting detection of genuine non-separability.  
For other values of $\alpha$ in which the local correlations are weaker, detection of genuine non-separability is allowed, as can be seen in the figure.

\section{Refined Tsirelson bounds on Mermin-Klyshko operators}\label{tighter_bounds_mk}
\subsection{Mermin-Klyshko operators}
The MK operators can be recursively defined, up to a normalization factor, as (Eq. (3) in \cite{collins2002bell}):
\begin{equation}\label{MermRecurs}
\begin{split}
    \mathcal{M}_N =  \frac{1}{2} \mathcal{M}_{N-1}  \left(A^{(N)}_0 + A^{(N)}_1\right) 
    \\
     + \frac{1}{2}\mathcal{M}_{N-1}^{'}\left(A^{(N)}_0 - A^{(N)}_1 \right) ,
\end{split}
\end{equation}
with the initial case of $\mathcal{M}_1 = A^{(1)}_0$. For consistency with the Svetlichny operators, we wish to normalize the MK operators such that the factor multiplying the correlation sums is always $1$. This requires to multiply the operators (as defined in Eq. \eqref{MermRecurs}) by $2^{N/2}$ for even $N$ and to multiply by $2^{(N-1)/2}$ for odd $N$.

We discuss separately the cases of even and odd $N$. For even $N$, since the MK operator is equivalent to the Svetlichny operator (as shown in Appendix \ref{BellProofs}), we may use our previous construction, yielding the same bounds. For odd $N$, the MK operator has $2^{N-1}$ elements, half the elements of the respective Svetlichny operator, which will require to change our construction accordingly.

\subsection{Refined Tsirelson bounds}
Consider a tripartite system under the bipartition $\underbrace{A^{(1)}}_X|\underbrace{A^{(2)}A^{(3)}}_Y$. Using inequality \eqref{MainIneqYX} combined with the triangle inequality we obtain (previously derived in \cite{carmi2018significance}, see Eq. (18) within):
\begin{widetext}
\begin{equation}\label{GenMermTri}
\begin{split}
\abs{\braket{\mathcal{M}_3}}\leq 
\abs{\braket{A^{(1)}_0A^{(2)}_0A^{(3)}_1}+\braket{A^{(1)}_0A^{(2)}_1A^{(3)}_0}} + \abs{\braket{A^{(1)}_1A^{(2)}_0A^{(3)}_0}-\braket{A^{(1)}_1A^{(2)}_1A^{(3)}_1}} \leq
\\
\leq  \sqrt{2 +  \braket{\{A^{(2)}_0A^{(3)}_1,A^{(2)}_1A^{(3)}_0\}}} + \sqrt{2 -  \braket{\{A^{(2)}_0A^{(3)}_0,A^{(2)}_1A^{(3)}_1\}}}.
\end{split}
\end{equation}
\end{widetext}
In contrast to our local bound on the Svetlichny operator, the bound here is comprised of bipartite correlations, since the elements $\braket{\{A^{(2)}_0A^{(3)}_1,A^{(2)}_1A^{(3)}_0\}}$ and $\braket{\{A^{(2)}_0A^{(3)}_0,A^{(2)}_1A^{(3)}_1\}}$ cannot be reduced to be local correlations.

It is impossible to recreate a similar result using inequality \eqref{MainIneqXY} under the same bipartition, $\underbrace{A^{(1)}}_X|\underbrace{A^{(2)}A^{(3)}}_Y$, as the MK operators for an odd number of parties do not include all possible permutations of the measured observables. By that we mean that when taking, for example, $\abs{\braket{A^{(1)}_iA^{(2)}_0A^{(3)}_0}+\braket{A^{(1)}_jA^{(2)}_0A^{(3)}_0}}$, there is no $i$ and $j$ such that both elements are in $\mathcal{M}_3$.

As such, our only bounds on $\mathcal{M}_3$ are in the form of Eq. \eqref{GenMermTri}, and similar bounds comprised of different combinations of the parties, had by using different by bipartitions. By the recursive definition of the MK operators, our bounds on $\abs{\braket{\mathcal{M}_{N}}}$ are four times the bounds on $\abs{\braket{\mathcal{M}_{N-2}}}$ and we have the following result for every $N$-party MK operator in the quantum regime, where $N$ is odd,
  \begin{equation}\label{BoundsMermGen}
\abs{\braket{\mathcal{M}_N}}
\leq  2^{N-3}\left( \sqrt{2 +\chi^{(n,m)}_+  } + \sqrt{2 -  \chi^{(n,m)}_-}\right),
\end{equation}
wherein $\chi^{(n,m)}_+ \defeq \braket{\{A^{(n)}_0A^{(m)}_1,A^{(n)}_1A^{(m)}_0\}}$ and 
$\chi^{(n,m)}_- \defeq \braket{\{A^{(n)}_0A^{(m)}_0,A^{(n)}_1A^{(m)}_1\}}$, for $n,m\in\{1,2,\ldots,N\}$ such that $n \neq m$. Note that $-2 \leq \chi^{(n,m)}_{\pm} \leq 2$.
 
 From our bounds, the MK operators, comprised of multipartite correlations between all the parties, are bounded by interactions between each two parties. The consequence of which is again a complementarity relation limiting the detection of non-full-separability.
 
 The maximal value of our bounds is $2^{N-1}$, the known quantum-algebraic bound, and is reached for $\chi^{(n,m)}_+  = -\chi^{(n,m)}_-  = 2$. The minimal value of our bounds is $0$, and is reached for $\chi^{(n,m)}_+  = -\chi^{(n,m)}_-  = -2$. The fact that our bounds are comprised of bipartite correlations, and can reach $0$, arises from our attempt to bound the MK operators, which do not include all possible permutations of the measured observables (for odd $N$), and is known to limit their ability to differentiate between various models underlying the correlations \cite{collins2002bell}. It follows from our bound that bipartite correlations within the multipartite system can prohibit the MK operators from crossing the classical bounds, and thus non-full-separability cannot be detected at all.

For cases in which our bounds are strictly tighter than the quantum-algebraic bounds, it is no longer the case that the algebraic and Tsirelson bounds are the same and the MK operators can now differentiate between algebraic (post-quantum \cite{popescu1994quantum}) and quantum theories.

\subsection{Classical bound - Odd number of parties}

Assume that parties $A^{(n)}$ and $A^{(m)}$ only share classical correlations (they are described by a separable LHV model). Thus, their operators commute and we have that $\chi^{(n,m)}_+  = \chi^{(n,m)}_-  = 2\braket{A^{(n)}_0A^{(n)}_1A^{(m)}_0A^{(m)}_1}$. As such, if just two parties, say $A^{(n)}$ and $A^{(m)}$, exhibit local correlations, the bounds from Eq. \eqref{BoundsMermGen} are now tighter, as expected:
 \begin{equation}
\abs{\braket{\mathcal{M}_N}}
\leq   2^{N-2}\sqrt{1+\sqrt{1-\left( \braket{A^{(n)}_0A^{(n)}_1A^{(m)}_0A^{(m)}_1} \right)^2}},
\end{equation}
in which we used the transition from Eq. \eqref{TransitionEq}. These bounds are constructed as our bounds on the Svetlichny operators and have a maximal value of $2^{N-2}\sqrt{2}$ and a minimal value of $2^{N-2}$. This accords well with the $2|1QM$ bound found in \cite{collins2002bell} for a tripartite system.

\section{Example - The Mermin-Klyshko operator}\label{MermExample}
Concepts similar to the examples shown for the Svetlichny operators hold for the MK operators. As such, we focus in this section on an example which is unique to the MK operators, testing our refined bounds on the MK operator $\mathcal{M}_3$, whilst still using the GHZ state and the operators defined in section \ref{SvetExample}.  

\begin{figure}[ht]
\includegraphics[scale=0.53]{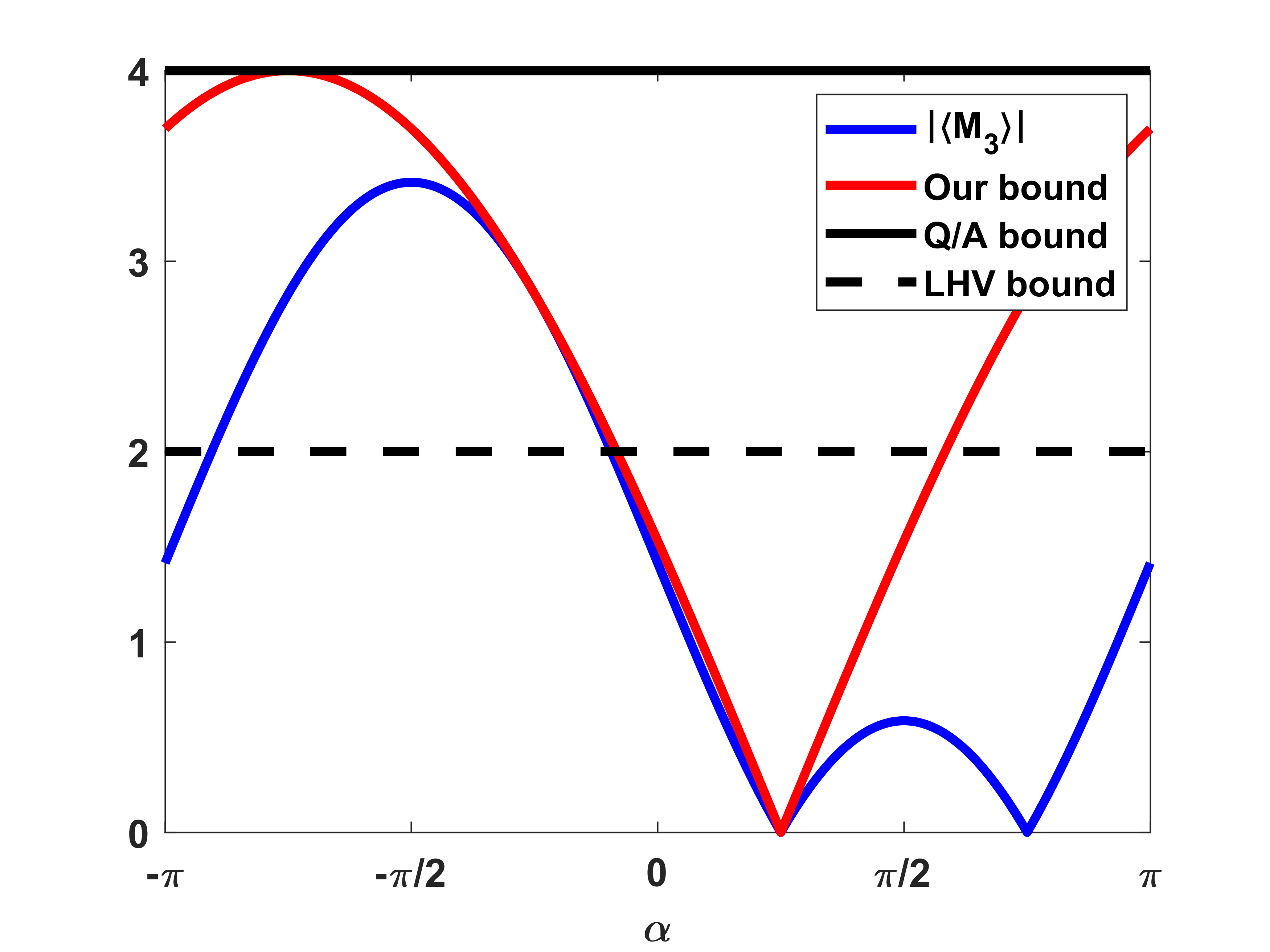}
\caption{\textbf{Demonstrating our bounds from Eq. \eqref{BoundsMermGen}.} Plotted are the MK operator $\mathcal{M}_3$ and our bound, calculated for a tripartite GHZ state. The quantum-algebraic (Q/A) bound is $4$ in the tripartite case. The complementarity between bipartite and nonlocal correlations is demonstrated: for every value of $\alpha$ in which our bound is smaller than $2$, non-full-separability cannot be detected.}
\centering
\label{MermZeroCorrFig}
\end{figure}

We again change the phase of the operators using the parameter $\alpha$: $(\theta_0^{(1)},\theta_0^{(2)},\theta_0^{(3)}) = (\alpha,0,0)$, $(\theta_1^{(1)},\theta_1^{(2)},\theta_1^{(3)}) = (-\pi/4,\pi/2,\pi/2)$, plotting our results in Fig. \ref{MermZeroCorrFig}. Our bound, plotted in red, is given by $\sqrt{2 +\chi^{(1,2)}_+  } + \sqrt{2 -  \chi^{(1,2)}_-}$ (Eq. \eqref{BoundsMermGen}) and we can see that for $\alpha = \pi/4$ it is zero, meaning that the MK operator must also be zero. As the classical bound (obtained in an LHV model factorizable for every party) is $2$, for any value of $\alpha$ in which our bound is smaller than $2$ non-full-separability cannot be detected. Furthermore, our bound is strictly tighter than the quantum-algebraic bound almost everywhere, making it is possible to differentiate between quantum and post-quantum models.

\section{Algebraic bounds on the Svetlichny operators}\label{algebraic_bounds}
Let us observe Eq. \eqref{MainIneqXY}. Its right hand side is independent of $Y_k$ and our choice to measure it. We may require that the right hand side would depend on our choice to measure $Y_k$, noting this dependence by $\braket{\{X_i,X_j\}}|_{Y_k}$. From the principle of relativistic independence \cite{carmi2019relativistic}, this sort of dependence is not possible, meaning that $\braket{\{X_i,X_j\}}|_{Y_k} = \braket{\{X_i,X_j\}}$. Neglecting relativistic independence and using the dependence on $Y_k$ gives rise to a new degree of freedom in our bounds on the Svetlichny inequalities. For example, Eq. \eqref{S2MinusBound1} can now be written as,
\begin{equation}
\begin{split}
\abs{\braket{\mathcal{S}_2^-}} \leq \sqrt{ 2 + \braket{\{A^{(1)}_0,A^{(1)}_1\}}|_{A^{(2)}_0}} \\
+ \sqrt{ 2 - \braket{\{A^{(1)}_0,A^{(1)}_1\}}|_{A^{(2)}_1}}.
\end{split}
\end{equation}
The maximal value of this bound can in principle be $4$, the algebraic bound, and is obtained when $\braket{\{A_0,A_1\}}_{A^{(2)}_0} = - \braket{\{A_0,A_1\}}_{A^{(2)}_1} = 2$. Using a recursive argument, our bounds on the Svetlichny operators for all $N$ can reach the algebraic bound, when neglecting the relativistic independence condition.

\section{Conclusions}\label{Conclusions}
In this work, we set out to examine Bell inequalities in multipartite systems, focusing on refinements of the well known Svetlichny and MK inequalities. Leveraging a fundamental constraint, akin to quantum uncertainty principles, namely the requirement that the covariance matrix must be positive semidefinite, we derived refined Tsirelson bounds for the Svetlichny and MK inequalities, applicable to any quantum state (pure or mixed) belonging to any finite-dimensional Hilbert space. Unlike the original inequalities, our bounds are not constant but depend on correlations in specific subsystems, signifying the interplay between multipartite nonlocal correlations and lower-order correlations (local or bipartite correlations depending on the specific inequality), as well as making it easier to detect correlations originating from post-quantum models.

Our bounds on the Svetlichny operator are comprised of strictly local correlations, giving a complementary relation between our local bound and the nonlocal Svetlichny operator. Employing this complementarity relation, we show that if the local correlations of just one party are maximal then genuine non-separability cannot be detected; and only when the local correlations of \emph{all} the parties are zero then the known Tsirelson bound may be saturated. 

For an even number of parties, the MK operators are equivalent to the Svetlichny operators, and our previous analysis holds. For an odd number of parties this is not the case and the bounds take the form of bipartite correlations. This stems from our attempt to bound operators which do not include all possible permutations of the measured observables, which is known to limit their ability to differentiate between various models underlying the correlations \cite{collins2002bell}. Again, we derived and emphasized the significance of a complementary relation between the nonlocal MK operator and bipartite correlations, showing cases in which bipartite correlations may allow or prohibit the detection of non-full-separability. For an odd number of parties, the MK operators cannot differentiate between quantum and post-quantum models. Under our construction this is no longer the case as our (quantum) bounds may be tighter than the known quantum-algebraic bound.

Our complementarity relations bear similarity to quantum monogamy relations \cite{coffman2000distributed,dhar2017monogamy} and to the more general idea of quantum correlations being a (shareable) resource \cite{sapienza2019correlations,adesso2016measures,shahandeh2019quantum}, as they limit our detection of non-separability in the system by posing constraints on one or two parties, i.e., if local or bipartite correlations are too strong, then we are unable to detect non-separability, as the correlations (being a resource) were ``consumed''. Bell inequalities and monogamy relations were already used in tandem for the study of quantum correlations, especially in the area of quantum cryptography \cite{ekert1991quantum,scarani2001quantum,seevinck2010monogamy,pawlowski2010security,tomamichel2013monogamy,xiang2023multipartite}. This presents an opportunity to apply our results to further study many-body quantum systems, as well as various quantum technologies such as entanglement-based quantum key distribution, in which our refined bounds could prove useful.

\begin{acknowledgments}
This work was supported by the Pazy Foundation (grant number 49579), by the Israeli Ministry of Science and Technology (grant number 17812), and by the Israel Innovation Authority (grant number 73795).
\end{acknowledgments}

\appendix

\section{Derivation of inequalities from the covariance matrix}\label{CovarianceProofs}

We can derive various helpful inequalities using the covariance matrix, which is defined for Hermitian operators $\{O_i\}_{i=0}^{N-1}$ as follows \cite{carmi2018significance}:
\begin{equation}
    \mathcal{C} \defeq M- VV^{T},
\end{equation}
wherein $M_{ij} \defeq \frac{1}{2}\braket{\{O_i,O_j\}}$ is an $N \times N$ matrix and $V_i \defeq \braket{O_i}$ is a column vector of length $N$. 
The covariance matrix is positive semidefinite, $\mathcal{C}\succeq 0 $  and therefore:
\begin{equation}
    M \succeq VV^{T}.
\end{equation}
For $N=2$, we define the vector $u \defeq \left[1, (-1)^m \right]^{T}$ for some integer $m$ and obtain the scalar inequality:
\begin{equation}
    u^TMu \geq u^TVV^{T}u.
\end{equation}
Multiplying the matrices while assuming that the operators are limited to give measurement outcomes of $\pm 1$ yields:
\begin{equation}\label{ConIneq}
\abs{\braket{O_0} +(-1)^{m} \braket{O_1}} \leq  \sqrt{2 +(-1)^{m} \braket{\{O_0,O_1\}}}.
\end{equation}
Note that $-2\leq \braket{\{O_0,O_1\}}\leq 2 $.

\section{Refined Tsirelson bounds on Svetlichny operators - Expanded derivation} \label{BellProofs}

Recall Eq. \eqref{SvetRecur}. Using the triangle inequality we deduce that: 
\begin{equation}
\abs{\braket{\mathcal{S}_{N}^{\pm}}} \leq \abs{\braket{\mathcal{S}_{N-1}^{\pm} A_0^{(N)}}} + \abs{\braket{\mathcal{S}_{N-1}^{\mp} A_1^{(N)}}}.
\end{equation}
As the measurement outcomes are $\pm 1$, we also have that:
 \begin{equation}
\abs{\braket{\mathcal{S}_{N-1}^{\pm} A_{i}^{(N)}}} \leq \abs{\braket{\mathcal{S}_{N-1}^{\pm}}},
\end{equation}
for $i \in \{1,2\}$. Thus, the bounds on $\abs{\braket{\mathcal{S}_{3}^{\pm}}}$ are twice the bounds presented on $\abs{\braket{\mathcal{S}_{2}^{\pm}}}$ in Eqs. \eqref{S2MinusBound1} and \eqref{S2MinusBound2}. Note that we do not have a bound for $\abs{\braket{\mathcal{S}_{3}^{\pm}}}$ using the operators of $A^{(3)}$, but using the fact that the Svetlichny operators are permutation invariant (as they include all combinations of the measured observables), solves this and we recursively obtain the results in Eq. \eqref{BoundsSvetGen}.

 It is true that the bound could have included a combination of different parties, for example:
\begin{equation}
\abs{\braket{\mathcal{S}_3^{\pm}}} \leq \sqrt{1+\sqrt{1-\eta^{(1)}}} + \sqrt{1+\sqrt{1-\eta^{(2)}}},
\end{equation}
but if we assume, without loss of generality, that $\eta^{(1)} \leq \eta^{(2)}$, then the single-party expression $2\sqrt{1+\sqrt{1-\eta^{(1)}}}$, which appears in our construction, is a tighter bound.

\section{Equivalence of the Svetlichny and MK operators}
Using Eq. \eqref{MermRecurs} we have (ignoring the constant factors multiplying the sums of the correlations):
    \begin{equation}
    \begin{split}
    \mathcal{M}_{N-1} =  \mathcal{M}_{N-2} \left(A^{(N-1)}_0 + A^{(N-1)}_1\right) \\
    + \mathcal{M}_{N-2}^{'}  \left(A^{(N-1)}_0 - A^{(N-1)}_1\right).
    \end{split}
    \end{equation}
We now define $Y \defeq A^{(N-1)} A^{(N)}$ and using binary encoding, e.g., $Y_1 = A^{(N-1)}_0 A^{(N)}_1$, we have:
\begin{equation}\label{MKN-2}
\begin{split}
     \mathcal{M}_N &=  \mathcal{M}_{N-1} \ \left(A^{(N)}_0 + A^{(N)}_1\right) + \mathcal{M}_{N-1}^{'} \ \left(A^{(N)}_0 - A^{(N)}_1\right) 
    \\
     & = \mathcal{M}_{N-2} \ \left( Y_0 + Y_1 + Y_2 + Y_3\right) \\
     &+\mathcal{M}_{N-2}^{'}\left( Y_0 + Y_1 - Y_2 - Y_3\right) 
     \\
     & + \mathcal{M}_{N-2}^{'}\left( Y_0 - Y_1 + Y_2 - Y_3\right) \\
     &+ \mathcal{M}_{N-2} \ \left( -Y_0 + Y_1 + Y_2 - Y_3\right)  =
     \\
     & = \mathcal{M}_{N-2} \ \left( Y_1 + Y_2 \right) + \mathcal{M}_{N-2}^{'} \ \left( Y_0 - Y_3 \right),
\end{split}
\end{equation}
where we have again ignored any constant factors. This result can also be obtained using Eq. (6) in \cite{collins2002bell}.

Using Eq. \eqref{SvetRecur},
\begin{equation}
   \mathcal{S}_{N-1}^{\pm} = \mathcal{S}_{N-2}^{\pm}\ A_0^{(N-1)} \mp \mathcal{S}_{N-2}^{\mp}\ A_1^{(N-1)},
\end{equation} 
which gives:
\begin{equation}
\begin{split}
&\mathcal{S}_{N}^{\pm} = \mathcal{S}_{N-1}^{\pm}\ A_0^{(N)} \mp \mathcal{S}_{N-1}^{\mp}\ A_1^{(N)}   
    \\
   & = \mathcal{S}_{N-2}^{\pm}\ Y_0 \mp \mathcal{S}_{N-2}^{\mp}\ Y_2 \mp \left( \mathcal{S}_{N-2}^{\mp}\ Y_1 \pm \mathcal{S}_{N-2}^{\pm}\ Y_3 \right) 
   \\
   & =\mathcal{S}_{N-2}^{\pm} \left( Y_0 - Y_3 \right) \mp \mathcal{S}_{N-2}^{\mp}\left( Y_1 + Y_2 \right).
\end{split}
\end{equation} 
As $N$ is even, $\abs{\mathcal{S}_{N-2}^{+}} = \abs{\left( \mathcal{S}_{N-2}^{-} \right)^{'}}$, a property noted in \cite{seevinck2002bell}. If $\mathcal{S}_{N-2}^{+} = -\left( \mathcal{S}_{N-2}^{-} \right)^{'}$, which happens for example in $\mathcal{S}_{2}^{+} = -\left( \mathcal{S}_{2}^{-} \right)^{'}$, we get $\mathcal{S}_{N}^{+} = \left( \mathcal{S}_{N}^{-} \right)^{'}$. If on the other hand $\mathcal{S}_{N-2}^{+} = \left( \mathcal{S}_{N-2}^{-} \right)^{'}$, we would get in a similar manner $\mathcal{S}_{N}^{+} = - \left( \mathcal{S}_{2}^{-} \right)^{'}$. As such, the sign of the equivalence between $\mathcal{S}_{N}^{+}$ and  $\mathcal{S}_{N}^{-}$ changes for each (even) $N$. 

For the first case of $N=2$, we have $-\left( \mathcal{S}_{2}^{+} \right)^{'} = \mathcal{S}_{2}^{-} = \mathcal{M}_2$, which gives:
\begin{equation}
\begin{split}
\mathcal{S}_{4}^{+} &= \mathcal{S}_{2}^{+} \left( Y_0 - Y_3 \right) - \mathcal{S}_{2}^{-}\left( Y_1 + Y_2 \right) \\ 
&= -\left(\mathcal{S}_{2}^{-} \right)^{'} \left( Y_0 - Y_3 \right) - \mathcal{S}_{2}^{-}\left( Y_1 + Y_2 \right) 
\\
& = -\mathcal{M}_2^{'} \left( Y_0 - Y_3 \right) - \mathcal{M}_2\left( Y_1 + Y_2 \right) = -\mathcal{M}_4.
\end{split}
\end{equation}

Now we have:
\begin{equation}\label{S4proof}
\begin{split}
\mathcal{S}_{6}^{-} &= \mathcal{S}_{4}^{-} \left( Y_0 - Y_3 \right) + \mathcal{S}_{N-2}^{+}\left( Y_1 + Y_2 \right) \\
&= \left(\mathcal{S}_{4}^{+}\right)^{'} \left( Y_0 - Y_3 \right) + \mathcal{S}_{N-2}^{+}\left( Y_1 + Y_2 \right) 
\\
&  = -\mathcal{M}_4^{'} \left( Y_0 - Y_3 \right) - \mathcal{M}_4\left( Y_1 + Y_2 \right) = -\mathcal{M}_6.
\end{split}
\end{equation}
The next case will give Eq. \eqref{S4proof} but with a minus sign and so on.

\section{Refined Tsirelson bounds on Mermin-Klyshko operators  - Expanded derivation}
Recall the bound found in Eq. \eqref{GenMermTri}, we may draw similar bound using the same process (using inequality \eqref{MainIneqYX} combined with the triangle inequality) but under different bipartitions. Under the bipartition
$\underbrace{A^{(2)}}_X|\underbrace{A^{(1)}A^{(3)}}_Y$ we get:
\begin{equation}
\begin{split}
\abs{\braket{\mathcal{M}_3}} &
\leq  \sqrt{2 +  \braket{\{A^{(1)}_0A^{(3)}_1,A^{(1)}_1A^{(3)}_0\}}}\\
&+ \sqrt{2 -  \braket{\{A^{(1)}_0A^{(3)}_0,A^{(1)}_1A^{(3)}_1\}}},
\end{split}
\end{equation}
and under the bipartition $\underbrace{A^{(3)}}_X|\underbrace{A^{(1)}A^{(2)}}_Y$ we get:
\begin{equation}
\begin{split}
\abs{\braket{\mathcal{M}_3}} &
\leq  \sqrt{2 +  \braket{\{A^{(1)}_0A^{(2)}_1,A^{(1)}_1A^{(2)}_0\}}}\\
&+ \sqrt{2 -  \braket{\{A^{(1)}_0A^{(2)}_0,A^{(1)}_1A^{(2)}_1\}}}.
\end{split}
\end{equation}

Using Eq. \eqref{MKN-2} and the triangle inequality we deuce that:
\begin{equation}
\begin{split}
\abs{\braket{\mathcal{M}_N}} &\leq \abs{\braket{\mathcal{M}_{N-2} A^{(N-1)}_0 A^{(N)}_1}} + \abs{\braket{\mathcal{M}_{N-2}A^{(N-1)}_1  A^{(N)}_0}}  \\
&+\abs{\braket{\mathcal{M}_{N-2}^{'} A^{(N-1)}_0 A^{(N)}_0}} + \abs{\braket{\mathcal{M}_{N-2}^{'} A^{(N-1)}_1 A^{(N)}_1}}.
\end{split}
\end{equation}
As the measurement outcomes are $\pm 1$, the bounds on $\abs{\braket{\mathcal{M}_5}}$ are four times the bounds presented on $\abs{\braket{\mathcal{M}_3}}$ in Eq. \eqref{GenMermTri} and its variations given by taking different bipartitions. Finally, by noting that the MK operators are permutation invariant (easily seen by Eq. \eqref{MKN-2} when recalling that $\mathcal{M}_3$ is also permutation invariant) we recursively get the results in Eq. \eqref{BoundsMermGen}.

It is true that the bound could have included a combination of different parties, for example:
\begin{equation}
\begin{split}
\abs{\braket{\mathcal{M}_5}} &\leq \sqrt{2 +\chi^{(1,2)}_+  } + \sqrt{2 -  \chi^{(1,2)}_-} \\
&+\sqrt{2 +\chi^{(1,3)}_+  } + \sqrt{2 -  \chi^{(1,3)}_-},
\end{split}
\end{equation}
but if we assume, without loss of generality, that $ \sqrt{2 +\chi^{(1,2)}_+  } + \sqrt{2 -  \chi^{(1,2)}_-} \leq  \sqrt{2 +\chi^{(1,3)}_+  } + \sqrt{2 -  \chi^{(1,3)}_-}$, then the single-party expression $2 \left( \sqrt{2 +\chi^{(1,2)}_+  } + \sqrt{2 -  \chi^{(1,2)}_-}\right)$, which appears in our construction, is a tighter bound.

\bibliography{ArticleReview}

\end{document}